\documentclass[letter,longauth,traditabstract]{aa} 

\usepackage{amsmath}
\usepackage{txfonts}
\usepackage{graphicx}
\usepackage{natbib}
\bibpunct{(}{)}{;}{a}{}{,} 
\pdfoutput=1
\begin{document}

\title{The \emph{Herschel} Virgo Cluster Survey: VII. Dust in cluster dwarf elliptical galaxies\thanks{\emph{Herschel} is an ESA space observatory with science
    instruments provided by European-led Principal Investigator
    consortia and with important participation from NASA.}}

\author{
I. De Looze\inst{1}
\and
M. Baes\inst{1} 
\and
S. Zibetti\inst{17}
\and
J. Fritz\inst{1}
\and
L. Cortese\inst{7}
\and 
J. I. Davies\inst{7}
\and
J. Verstappen\inst{1}
\and 
G.~J. Bendo\inst{2}
\and
S. Bianchi\inst{3}
\and
M. Clemens\inst{6}
\and 
D. J. Bomans\inst{4}
\and
A. Boselli\inst{5}
\and
E. Corbelli\inst{3}
\and
A. Dariush\inst{7}
\and
S. di Serego Alighieri\inst{3}
\and
D. Fadda\inst{8}
\and
D. A. Garcia-Appadoo\inst{9}
\and
G. Gavazzi\inst{10}
\and
C. Giovanardi\inst{3}
\and
M. Grossi\inst{11}
\and
T. M. Hughes\inst{7}
\and
L. K. Hunt\inst{3}
\and
A. P. Jones\inst{12}
\and
S. Madden\inst{13}
\and
D. Pierini\inst{18}
\and
M. Pohlen\inst{7}
\and 
S. Sabatini\inst{14}
\and
M. W. L. Smith\inst{7}
\and
C. Vlahakis\inst{15} 
\and
E. M. Xilouris\inst{16}
}

\institute{
Sterrenkundig Observatorium, Universiteit Gent, Krijgslaan 281 S9, B-9000 Gent, Belgium 
\and
Astrophysics Group, Imperial College London, Blackett Laboratory, Prince Consort Road, London SW7 2AZ, UK 
\and
INAF-Osservatorio Astrofisico di Arcetri, Largo Enrico Fermi 5, 50125 Firenze, Italy 
\and
Astronomical Institute, Ruhr-University Bochum, Universitaetsstr. 150, 44780 Bochum, Germany 
\and
Laboratoire d'Astrophysique de Marseille, UMR 6110 CNRS, 38 rue F. Joliot-Curie, F-13388 Marseille, France 
\and
INAF-Osservatorio Astronomico di Padova, Vicolo dell'Osservatorio 5, 35122 Padova, Italy
\and
Department of Physics and Astronomy, Cardiff University, The Parade, Cardiff, CF24 3AA, UK
\and
NASA \emph{Herschel} Science Center, California Institute of Technology, MS 100-22, Pasadena, CA 91125, USA 
\and
European Southern Observatory, Alonso de Cordova 3107, Vitacura, Santiago, Chile 
\and
Universita' di Milano-Bicocca, piazza della Scienza 3, 20100, Milano, Italy 
\and
CAAUL, Observat\'orio Astron\'omico de Lisboa, Universidade de Lisboa,
Tapada da Ajuda, 1349-018, Lisboa, Portugal
\and
Institut d'Astrophysique Spatiale (IAS), Batiment 121, Universite Paris-Sud 11 and CNRS, F-91405 Orsay, France 
\and
Laboratoire AIM, CEA/DSM- CNRS - Universit\'e Paris Diderot, Irfu/Service d'Astrophysique, 91191 Gif sur Yvette, France 
\and
INAF-Istituto di Astrofisica Spaziale e Fisica Cosmica, via Fosso del Cavaliere 100, I-00133, Roma, Italy 
\and
Leiden Observatory, Leiden University, P.O. Box 9513, NL-2300 RA Leiden, The Netherlands 
\and
National Observatory of Athens, I. Metaxa and Vas. Pavlou, P. Penteli, GR-15236 Athens, Greece 
\and
Max-Planck-Institut fuer Astronomie, Koenigstuhl 17, D-69117
Heidelberg,  Germany 
\and
Max-Planck-Institut fuer extraterrestrische Physik, Giessenbachstrasse, 85748 Garching, Germany
}

\date{\today}

\abstract{
We use the Science Demonstration Phase data of the \emph{Herschel} Virgo Cluster Survey to search for dust emission of early-type dwarf galaxies in the central regions of the Virgo Cluster as an alternative way of identifying the interstellar medium. 
We present the first possible far-infrared detection of cluster early-type dwarf galaxies: \object{VCC\,781} and \object{VCC\,951} are detected at the $10\sigma$ level in the SPIRE 250~$\mu$m image. Both detected galaxies have dust masses of the order of $10^5$~M$_\odot$ and average dust temperatures $\approx$ 20K.  
The detection rate (less than 1\%) is quite high compared to the 1.7$\%$ detection rate for H{\sc{i}} emission, considering that dwarfs in the central regions are more H{\sc{i}} deficient. We conclude that the removal of interstellar dust from dwarf galaxies resulting from ram pressure stripping, harassment, or tidal effects must be as efficient as the removal of interstellar gas.}


\maketitle

\section{Introduction}

Early-type dwarf galaxies (dEs) are the dominant morphological type in galaxy clusters. They were originally seen as a rather homogeneous population of dwarf galaxies with an old stellar age, no features resembling any recent or ongoing star formation, and no indications of a significant interstellar medium (ISM). This viewpoint has changed radically in the past few years. Deep imaging observations of dEs have revealed a heterogeneous morphology. In particular, the population of dEs can be subdivided into nucleated and non-nucleated subclasses and several papers report evidence of disks, spiral structure, bars or star formation \citep[e.g.,][]{Jerjen2000, Barazza2002, Geha2003, Graham2003, GrahamGuzman2003, DeRijcke2003, Lisker2006_1, Lisker2006_2}. Kinematical studies demonstrate that dEs are not simple pressure-supported systems: some dEs seem to be rotationally supported \citep{Simien2002, Pedraz2002, Geha2003, DeRijcke2003, VanZee2004, Toloba2009}, whereas others show evidence of kinematically decoupled cores \citep{2004A&A...426...53D, 2006A&A...445L..19T}.

Adding to this morphological and kinematical inhomogeneity is the detection of a significant interstellar medium in various dEs. Atomic and molecular gas was discovered for the first time in the Local Group dEs NGC\,185 and NGC\,205 using deep VLA and BIMA observations \citep{1997ApJ...476..127Y}.
In the Virgo cluster, \citet{Conselice2003} estimate a H{\sc{i}} detection rate of 15\% for dEs down to a 3$\sigma$ limit of  $8 \times 10^{6}$~M$_\odot$, while
\citet{Alighieri2007} report a 1.7$\%$ H{\sc{i}} detection rate with a 3$\sigma$ detection limit of $3.5 \times 10^{7}$~M$_\odot$. Large amounts of cold interstellar matter are unexpected in dEs, as both internal (supernova explosions) and external effects (ram-pressure stripping, galaxy interactions, tidal effects) are thought to be able to expel the gas from the shallow potential on short timescales \citep{2004ANS...325..122M, VanZee2004, 2005A&A...433..875R, 2008ApJ...674..742B, 2008A&A...489.1015B, Valcke2008}. That the H{\sc{i}} detected dEs are found preferentially near the edge of the cluster supports the idea that understanding environmental effects is crucial to constraining the evolutionary history of dEs. \citet{Buyle2005} and \citet{Bouchard2005} performed H{\sc{i}} observations in the outskirts of the Fornax cluster and the Sculptor group, respectively, and confirmed the tendency of gas-deficient galaxies to be located near the center of the cluster.

Continuum emission from interstellar dust is a promising alternative way to determine the ISM content of dEs. In particular, the ISM in the immediate surroundings of M87 can be traced by dust emission, as strong radio continuum emission considerably reduces the ALFALFA H{\sc{i}} detection sensitivity within 1$\degr$ of M87 \citep{2007AJ....133.2569G}. To date, the Andromeda satellites NGC\,205 and NGC\,185 are the only dEs that have been detected in the far-infrared \citep{Haas1998, Marleau2006, Marleau2009}. In particular, no dust emission has been detected from cluster dEs. While extinction features in optical images are indicative of dust in at least some cluster dEs \citep{Ferrarese2006, Lisker2006_1, Lisker2006_2}, the direct detection of dust emission has been hampered by the limited resolution and sensitivity of the previous generation of far-infrared instrumentation. The advent of the \emph{Herschel} Space Observatory \citep{Herschel} offers new possibilities for mapping the ISM in dEs. In this paper, we report on our search for far-infrared emission from dEs in the Virgo Cluster, based on the Science Demonstration Phase (SDP) observations of the \emph{Herschel} Virgo Cluster Survey (HeViCS\footnote{More details on HeViCS can be found on http://www.hevics.org.}). In Sect.~{\ref{Observations.sec}}, we describe the observations, data reduction, and sample selection. In Sect.~{\ref{Detections.sec}}, we present our detections and in Sect.~{\ref{Stacking.sec}} we discuss the results of a stacking analysis of the non-detections. Sect.~{\ref{Discussion.sec}} gives our summary.

\begin{figure*}
  \centering
  \includegraphics[width=\textwidth]{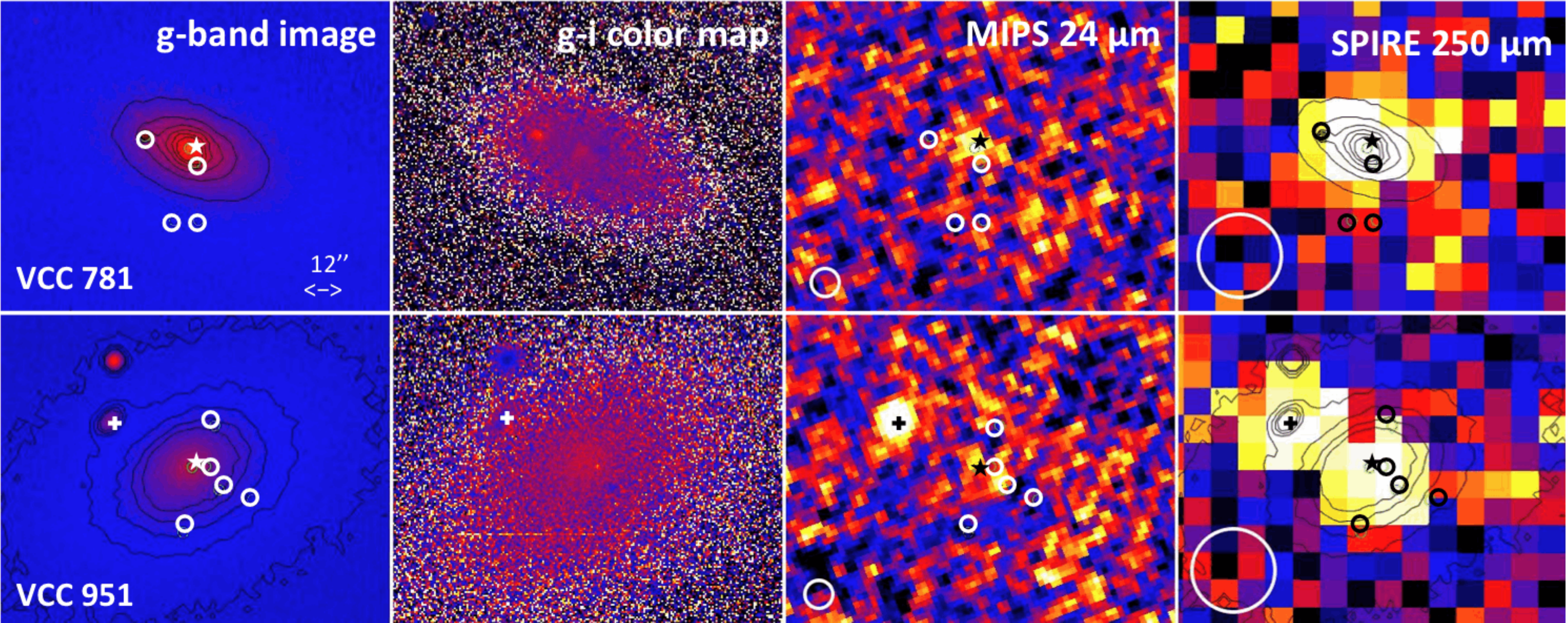}
  \caption{
All images are on the same scale. The FWHM of the beam is denoted in the MIPS 24 $\mu$m and SPIRE 250 $\mu$m images. g-band contours are overlaid on the SPIRE 250 $\mu$m image. The dwarfs position is indicated as a star, SDSS background sources as open circles, and the neighbouring source of VCC\,951 as a cross.}
  \label{VCC781.pdf}
\end{figure*}

\section{Observations and sample selection}
\label{Observations.sec}

The central $4\times4$~deg$^2$ region of the Virgo Cluster was observed by \emph{Herschel} on 29 November 2009 as part of the HeViCS SDP observations. The observations were performed in parallel scan map mode, which means that both PACS \citep{PACS} and SPIRE \citep{SPIRE} data were collected. The scan speed was 60\arcsec/s and the field was covered in nominal and orthogonal directions to minimize $1/f$ noise. The PACS and SPIRE data were reduced with HIPE, with reduction scripts based on the standard reduction pipelines. For more details of the HeViCS SDP data reduction, we refer to \citet{HeViCS-paper1}. 
We used the HeViCS SDP observations to search for dust emission from dEs in the central regions of the Virgo Cluster. Selecting the galaxies with morphological type dE and dS0 in the Goldmine catalogue \citep{Gavazzi2003} resulted in an initial sample of 370 dEs. After the removal of 16 galaxies classified as background sources and the exclusion of 115 sources with an additional optical source within a 6\arcsec\ radius, our final sample consisted of 239 dEs. Blind aperture photometry was applied to the PACS and SPIRE maps based on the positions of each of these 239 galaxies using Source Extractor \citep{1996A&AS..117..393B}.

\section{Detected dEs}
\label{Detections.sec}

From this sample of 239 sources, 11 dEs were detected above 3$\sigma$ in the SPIRE 250 $\mu$m image. Nevertheless, we only report the clear detection of 2 dEs,\object{VCC\,781} and \object{VCC\,951}, excluding all sources raising any kind of doubt. As such, VCC\,1502 and VCC\,788 were ruled out because their FIR location differs more than FWHM/2 from their optical position. VCC\,832 was excluded because SDSS images detect a background source with substructure in the residual image, possibly indicating the presence of dust.
Four other objects (VCC\,752, VCC\,815, VCC\,1272, VCC\,1512) fall within a crowded field of sources, enhancing the chances of their being background sources. For instance, \citet{Binggeli1985} reports the detection of a blue compact background galaxy, only $4\arcsec$ to the west of the nucleus of VCC\,815. Furthermore, VCC\,752 and VCC\,1272 appear to have SEDs that do not correspond to a single grey body. This failure in fitting a SED might either be caused by poor flux determination (the error bars are large compared to the low number of counts in the detection) or be a true indication that both detections originate in sources other than dEs.
The other two sources VCC\,1578 and VCC\,767 are clearly detected in the SPIRE 250 $\mu$m image at 4.8 and 4$\sigma$, respectively, but their high FIR flux relative to their faint optical appearance (both objects are low surface brightness [LSB] galaxies with B-band magnitudes of the order $\sim$21 mag) aroused suspicion. 

Dust has already been detected in LSBs \citep{Hinz2007}, but mainly in those with large amounts of star formation responsible for dust heating. A lack of evidence of recent star formation in VCC\,767 and VCC\,1578 strongly argues against the association of the 250 $\mu$m emission with the corresponding dEs. Furthermore, we can identify a background source in the immediate surroundings of VCC\,1578, which possibly contaminates the SPIRE 250 $\mu$m emission.

Nonetheless, these questionable detections may still be caused by dust emission from dEs. Deeper fields in the future might reveal dust emission in PACS images, where the higher spatial resolution may be able to rule out possible contamination by background sources.

On the other hand, \object{VCC\,781} and \object{VCC\,951} are detected unambiguously in the SPIRE 250~$\mu$m image at 9.8 and 11.1$\sigma$, respectively (see Fig.~\ref{VCC781.pdf}). Adopting a fair 2$\sigma$ detection cutoff in the other filters, \object{VCC\,781} is also detected in the PACS 100~$\mu$m, PACS 160~$\mu$m, SPIRE 350~$\mu$m, and SPIRE 500~$\mu$m bands.
\object{VCC\,951} has additional detections only in both SPIRE 350~$\mu$m and 500~$\mu$m images. Unfortunately, it is impossible to measure fluxes for \object{VCC\,951} at 350 $\mu$m and 500 $\mu$m because \object{VCC\,951} and its neighbouring source, SDSS J122655.56+114000.5, are not resolved at these longer wavelengths.

Since the SPIRE 250~$\mu$m image contains a vast number of sources, we need to address the possibility of false detections due to background sources. To perform this check accurately, we search all relevant images/catalogs (SDSS, Goldmine, 2MASS, UKIDSS, \emph{Spitzer}) for sources within 18.1 \arcsec (FWHM of SPIRE 250~$\mu$m) of the dwarf positions. For \object{VCC\,781}, this identifies four SDSS sources within this PSF region that may be contaminating the dwarf emission. The superior resolution in the PACS 160~$\mu$m image discards three of these optically identified objects, while the remaining source is rejected as a possible background candidate for the IR emission, based on the 24 $\mu$m image (data from Fadda et al. in prep.), which has an even higher spatial resolution, and is clearly extended. Hence, the coincidence of the position of detections at various wavelengths makes us confident that the IR emission originates in \object{VCC\,781}.

The SDSS DR7 catalog provides us with 5 detections within the PSF region of \object{VCC\,951}, three of which are very likely to be artifacts in the source extraction, since we detected them in neither the SDSS nor the UKIDSS images during our visual inspection. The remaining two, morphologically classified as stellar-like objects, are located at distances of 8'' and 10'', and are also detected in the UKIDSS images. The 24 $\mu$m image (see Fig.~\ref{VCC781.pdf}) shows some evidence of emission from both \object{VCC\,951} and one of the two aforementioned sources, which is spatially resolved at these wavelengths. Based on these data, we cannot definitely conclude that the IR emission detected by SPIRE comes from our dwarf galaxy, but deeper optical and NIR observations may help us to address the issue.

We conclude that, while our analysis is unable to definitely exclude a possible contamination from background sources, the weight of evidence is in favour of a true detection of dust in emission from dEs, and in particular from \object{VCC\,781}.

These detected dEs have remarkable morphologies. \citet{Binggeli1985} classifies \object{VCC\,781} as dS0$_{3}$(5),N and \object{VCC\,951} as dS0(2),N or dE2 pec,N. These galaxies are also found by \citet{Lisker2006_1} to harbor central substructures other than a disk, while \citet{Lisker2006_2} identified blue central colors in \object{VCC\,781} and \object{VCC\,951}. 
Figure~\ref{VCC781.pdf} presents the $g-i$ color images in which \object{VCC\,781} and \object{VCC\,951}, respectively, exhibit an obvious gradient in their radial $g-i$ color profiles, strengthening the assumption of recent star formation in the central regions \citep[see also][]{Lisker2006_2}. 

According to the classification criterion (3 $<$ FUV-H $<$ 6) adopted in \citet{2008ApJ...674..742B}, \object{VCC\,781} and \object{VCC\,951} can also be classified as possible transition objects. Although the FUV-H colours of 6.6 and 6.9 mag for \object{VCC\,781} and \object{VCC\,951}, respectively, do not satisfy this relation, both galaxies are clearly located at the blue end of the dE galaxies in \citet{2008ApJ...674..742B}. That SDSS nuclear spectra of both galaxies exhibit deep Hydrogen absorption lines (EW[H$\delta$] $>$ 4 \AA) indicates that they are in a post-starburst phase. This, and their 24 $\mu$m emission, which is indeed centered on the optical nuclei and concentrated ($\sim 6''$ for \object{VCC781} and point-like for \object{VCC951}), argues in favour of a connection between dust emission and a recent episode of star-formation.

To estimate the dust masses, we determined fluxes in bands where we had detections $>$ 2$\sigma$. Initial apertures were fixed in the SPIRE 250 $\mu$m images based on the flattening of the growth curve. These apertures were subsequently adjusted to the pixel scale in other bands such that all apertures cover the same physical area. 
We first determined a representative dust temperature using the assumption that the dust is in thermal equilibrium with the interstellar radiation field. We used the Monte Carlo code SKIRT, which was initially developed to investigate the effects of dust extinction on the photometry and kinematics of galaxies \citep{2003MNRAS.343.1081B}, but evolved into a flexible tool that can model the absorption, scattering, and thermal emission of circumstellar discs and dusty galaxies \citep[e.g][]{2007BaltA..16..101V, Baes}. The stellar body of each galaxy was represented as an exponential profile, with parameters taken from the Goldmine database. The dust was assumed to have the same distribution as the stars. For the intrinsic SED of the model, we used the elliptical galaxy template SED from the PEGASE library \citep{1997A&A...326..950F}. Adding these ingredients, we find a representative dust equilibrium temperature of 20.7~K and 19.4~K for \object{VCC\,781} and \object{VCC\,951}, respectively. Relying on these temperatures, we compute the dust masses according to
\begin{equation}
  M_{\text{d}}
  =
  \frac{S_{250}\,D^2}{\kappa_{250}\,B_{250}(T_{\text{d}})},
\end{equation}
where $D=16.5$~Mpc is the distance to the Virgo Cluster \citep{2007ApJ...655..144M}, $\kappa_{250}=0.4$~m$^2$~kg$^{-1}$ is the dust absorption coefficient \citep{2001ApJ...551..807D}, and $B_{250}(T_{\text{d}})$ is the Planck function for the modeled temperature $T_{\text{d}}$. We obtain dust masses of $1.85\times 10^5$~M$_\odot$ and $1.28\times 10^5$~M$_\odot$ for \object{VCC\,781} and \object{VCC\,951}, respectively.

\begin{figure}
 \centering
  \includegraphics[width=0.43\textwidth]{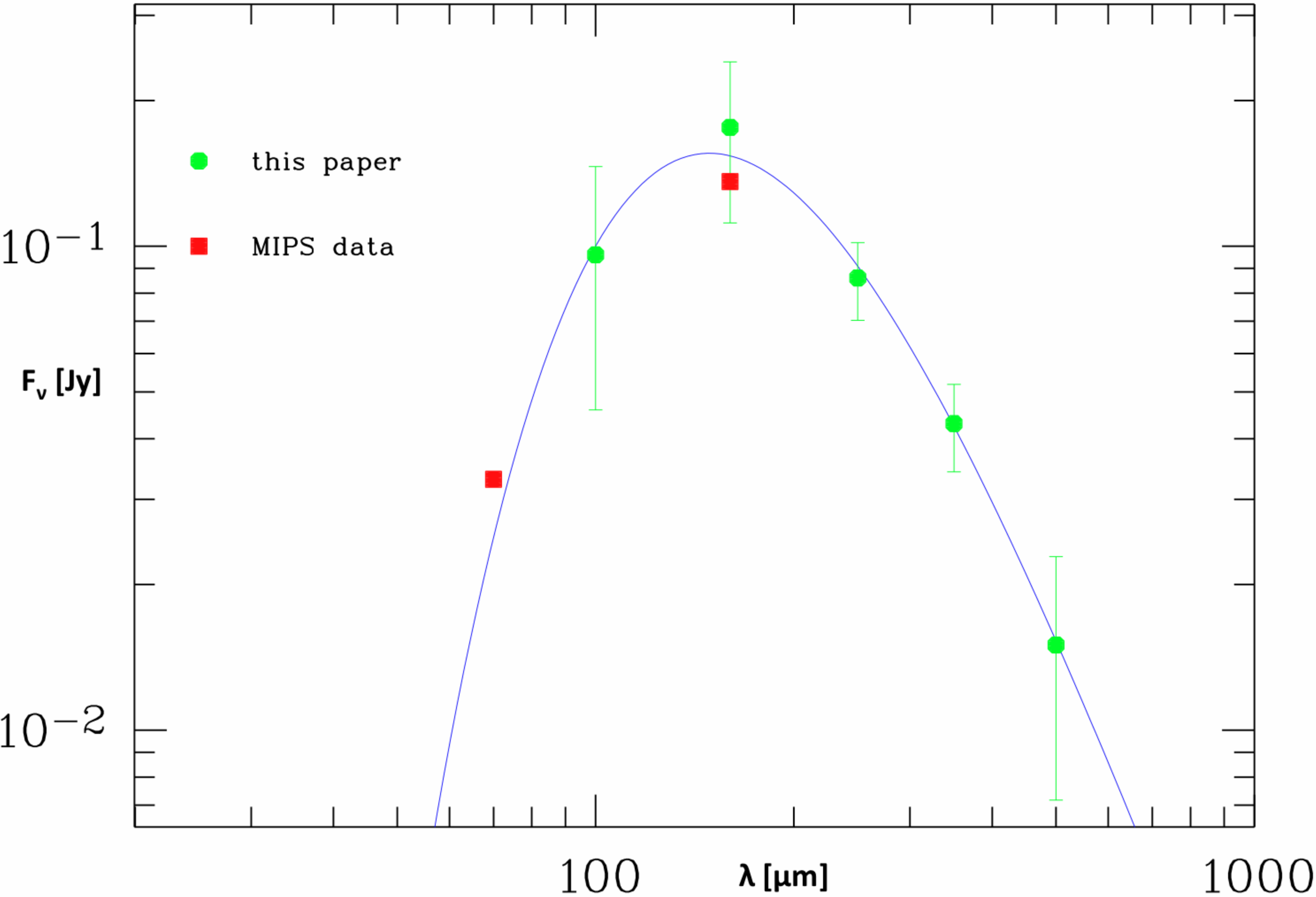}
  \caption{The best-fitting grey body for VCC\,781 with T$_{\text{d}}$ = 19.5 K and dust mass M$_{\text{d}}$ = $2.24 \times 10^{5}$~M$_{\odot}$.}
  \label{greybody.pdf}
\end{figure}

As a check for consistency with datapoints other than the SPIRE 250 $\mu$m flux, we determine a second estimate of the temperature and dust mass based on grey body fitting. A single grey body fit to the 5 PACS and SPIRE fluxes gives T$_{\text{d}}$ = 19.5 K and M$_{\text{d}}$ = $2.24 \times 10^{5}$~M$_{\odot}$ for \object{VCC\,781} (see Fig. \ref{greybody.pdf}). These results are fully consistent with the temperature and dust mass of \object{VCC\,781} derived above. The MIPS 70 and 160 $\mu$m datapoints also satisfy the fitted greybody curve well (see Fig.~\ref{greybody.pdf}).

Neither of the two dwarf galaxies is detected in H{\sc{i}}. An H{\sc{i}} mass upper limit was set to $2.3\times10^7~M_\odot$ for \object{VCC\,781} \citep{Gavazzi2003} and $8.0\times10^6~M_\odot$ for \object{VCC\,951} \citep{Conselice2003}. Combining the estimated dust masses with these H{\sc{i}} upper limits, and neglecting molecular gas, we find H{\sc{i}} gas-to-dust upper limits of 124.3 and 62.5, respectively. These values are much lower than the expected canonical gas-to-dust ratio, possibly indicating that there is a dependence on other factors such as the metallicity and the environment.

\section{Stacking analysis of non-detections}
\label{Stacking.sec}

Apart from the two detections described in the previous section, we performed a stacking analysis of non-detected sources. In a first step we masked all bright sources in the surroundings of the dwarf galaxy (we prevented the inner pixel from being masked). We enlarged these masks using a boxcar with smoothing kernel of 3 pixels to cover the whole source.  The final stacking algorithm weighs the images with the inverse of the background variance to acquire the mean value in each pixel (of all unmasked pixels).

Stacking the SPIRE 250 $\mu$m images of all 227 non-detected early-type dwarf galaxies results in no detection. When taking apertures of size equal to the SPIRE 250 $\mu$m beam at random positions and calculating the standard deviation in their distribution, we derive an estimated background rms of 0.38 mJy. Assuming an average dust temperature of T$_{\text{d}}$ = 20K, we convert this noise estimate to a 3$\sigma$ dust mass detection limit of M$_{\text{d}} = 2.44 \times 10^{3}$~M$_{\odot}$.

Since the noise measured locally has an average value of rms = 5 mJy/beam, the stacking is not significantly affected by confusion noise (the rms almost scales as Gaussian noise, 5/$\sqrt{227}$ = 0.33 $\approx$ 0.38). The complete survey of 4 cross-scans will reach a 1$\sigma$ sensitivity of 4.72 mJy on small spatial scales, considering both instrumental and confusion noise (the latter is determined to be 4 mJy at SPIRE 250 $\mu$m). 
Assuming that the stacking procedure is only marginally affected by confusion noise, this value translates into a 3$\sigma$ detection limit of M$_{\text{d}} = 2.0 \times 10^{3}$~M$_{\odot}$. Comparing both values, we conclude that the stacked image nearly reaches the sensitivity limit, as the detection limit will only improve by a factor of 1.2 after the complete survey.

\section{Discussion and conclusion}
\label{Discussion.sec}

We have reported on our search for far-infrared dust emission in a sample of 239 dEs at the centre of the Virgo Cluster. With a detection limit of about $10^4~M_\odot$, we have detected two objects at 10$\sigma$ that have a high probability of being true detections of dEs. These two dwarf galaxies,\object{VCC\,781} and \object{VCC\,951}, are the first detections of dust emission in dEs, apart from the Andromeda satellites NGC\,205 and NGC\,185. Applying a radiative transfer model to determine the equilibrium dust temperature in their interstellar radiation field, we determined temperatures of the order of 20~K in both galaxies. We estimate  dust masses of the order M$_{\text{d}} = 10^5$~M$_\odot$ in both dEs. 

In addition to independent detections of both atomic and molecular gas in dEs, the detection of dust emission here confirms the hypothesis that at least some dEs are transition objects gradually evolving from late-type to early-type, while falling into the cluster.

The detection of an interstellar dust medium in two dEs is rather unexpected, given the efficiency with which the ISM is removed from dwarf galaxies. \citet{2008ApJ...674..742B} found that gas removal by ram pressure stripping is extremely efficient in dwarfs galaxies, there being a decrease by $\sim$2 orders of magnitude in $\sim$150~Myr of time. In agreement with this rapid morphological transition, only a small number ($10-16 \%$) of dwarf galaxies exhibiting properties intermediate between star-forming and spheroidal galaxies are present in the Virgo Cluster. 

\citet{Alighieri2007} find an H{\sc{i}} detection rate in the Virgo cluster of only 1.7\% to a 3$\sigma$ mass limit M$_{\text{H{\sc{i}}}} = 3.5 \times 10^7$~M$_\odot$. Adopting a canonical gas-to-dust ratio of 1000 (which might still be a conservative lower limit as dEs in Virgo are expected to have metallicities significantly lower than observed in our own Milky Way), we recover a $3\sigma$ dust mass limit of $3.5 \times 10^4$~M$_\odot$, which is more or less the $3\sigma$ dust mass detection limit in the HeViCS SDP field. Our detection rate is also compatible with the 15\% found by \citet{Conselice2003}. Their H{\sc{i}} upper limit corresponds to a 3$\sigma$ dust mass limit of $8 \times 10^{3}$~M$_\odot$, well below the SDP detection limit. When also taking into account the position in the cluster, we consider our detection rate of less than 1\% surprisingly high. Indeed, H{\sc{i}} detected dEs are preferentially located in the outskirts of the cluster and may not have been dramatically affected so far by the intracluster medium and interactions with other galaxies \citep{Conselice2003, Buyle2005}. \citet{Alighieri2007} also report that the majority of the detected H{\sc{i}}-sources are located near the edges of the Virgo cluster. If gas and dust are tightly coupled and dust is as easily removed from the galaxy as gas by ram pressure stripping, harassment, or tidal effects, we expect that the dEs in the central regions of the Virgo Cluster will also be the most dust deficient. Clear evidence of efficient dust stripping in the Virgo Cluster is seen for the first time in H{\sc{i}}-deficient spiral galaxies \citep{Luca}. Future HeViCS observations will be able to provide an answer to this question -- the HeViCS survey will not only go deeper (each field will be covered 4 times), but also wider (the total area covered will be $\sim$64 deg$^2$). A comparison of the far-infrared detection rate with the position in the cluster will tell us whether environmental effects have any influence on the detection rate.


\begin{thebibliography}{}

\bibitem[Baes et al.(2003)]{2003MNRAS.343.1081B} Baes, M., et al.\ 2003, \mnras, 343, 1081

\bibitem[Baes et al.(2010)]{Baes} Baes, M., Fritz,~J., Gadotti,~D.~A. et al.\ 2010, A\&A, this issue

\bibitem[Barazza et al.(2002)]{Barazza2002} Barazza, F.~D., Binggeli, B. \& Jerjen, H.\ 2002, A\&A, 391, 823

\bibitem[Bertin 
\& Arnouts(1996)]{1996A&AS..117..393B} Bertin, E., \& Arnouts, S.\ 1996, \aaps, 117, 393 

\bibitem[Binggeli et al.(1985)]{Binggeli1985} Binggeli, B., Sandage, 
A., \& Tammann, G.~A.\ 1985, \aj, 90, 1681 
\bibitem[Boselli et al.(2008a)]{2008ApJ...674..742B} Boselli, A., Boissier, 
S., Cortese, L., \& Gavazzi, G.\ 2008, \apj, 674, 742 
\bibitem[Boselli et 
al.(2008b)]{2008A&A...489.1015B} Boselli, A., Boissier, S., Cortese, L., \& Gavazzi, G.\ 2008, \aap, 489, 1015 

\bibitem[Bouchard et al.(2005)]{Bouchard2005} Bouchard, A., et al.\ 2005, \aj, 130, 2058 
\bibitem[Buyle et al.(2005)]{Buyle2005} Buyle, P. et al.\ 2005, MNRAS, 360, 853
\bibitem[Conselice et al.(2003)]{Conselice2003} Conselice C.~J. et al. \ 2003, \apj, 591, 167 

\bibitem[Cortese et al.(2010)]{Luca} Cortese, L., Davies,~J.~I., Pohlen,~M. et al.\ 2010, A\&A, this issue

\bibitem[Davies et al.(2010)]{HeViCS-paper1} Davies, J.~I., Baes,~M., Bendo,~G.~J. et al.\
  2010, A\&A, this issue
\bibitem[De Rijcke et al.(2003)]{DeRijcke2003} De Rijcke S. et al. \ 2003, A\&A, 400, 119

\bibitem[De Rijcke et 
al.(2004)]{2004A&A...426...53D} De Rijcke, S., Dejonghe, H., Zeilinger, W.~W., \& Hau, G.~K.~T.\ 2004, \aap, 426, 53 
\bibitem[di Serego Alighieri et 
al.(2007)]{Alighieri2007} di Serego Alighieri, S., et al.\ 2007, \aap, 474, 851 
\bibitem[Draine \& Li(2001)]{2001ApJ...551..807D} Draine, B.~T., \& Li, A.\ 2001, \apj, 551, 807 
\bibitem[Ferrarese et al.(2006)]{Ferrarese2006} Ferrarese, L. et al.\ 2006 \apjs, 164, 334

\bibitem[Fioc \& Rocca-Volmerange(1997)]{1997A&A...326..950F} Fioc, M., \& Rocca-Volmerange, B.\ 1997, \aap, 326, 950

\bibitem[Gavazzi et al. (2003)]{Gavazzi2003} Gavazzi G. et al. \ 2003, A\&A, 400, 451
\bibitem[Geha et al. (2003)]{Geha2003} Geha, M., Guhathakurta, P. \& van der Marel, R.~P. \ 2003, \aj, 126, 1794
\bibitem[Giovanelli et al.(2007)]{2007AJ....133.2569G} Giovanelli, R., et 
al.\ 2007, \aj, 133, 2569 
\bibitem[Graham et al.(2003a)]{Graham2003} Graham, A.~W. et al.\ 2003, \aj, 125, 2951
\bibitem[Graham et al.(2003b)]{GrahamGuzman2003} Graham, A.~W. \& Guzman, R.\ 2003, \aj, 125, 2936
\bibitem[Griffin et al.(2010)]{SPIRE} Griffin, M., et al.\ 2010, A\&A, this issue
\bibitem[Haas(1998)]{Haas1998} Haas, M.\ 1998, \aap, 337, L1
\bibitem[Hinz et al.(2007)]{Hinz2007} Hinz, J.~L., Rieke, M.~J., 
Rieke, G.~H., Willmer, C.~N.~A., Misselt, K., Engelbracht, C.~W., Blaylock, 
M., \& Pickering, T.~E.\ 2007, \apj, 663, 895
\bibitem[Jerjen et al.(2000)]{Jerjen2000} Jerjen, H., Kalnajs, A., \& Binggeli, B.\ 2000, A\&A, 358, 845
\bibitem[Lisker et al.(2006a)]{Lisker2006_1} Lisker, T., Grebel, E.~K.,
 \& Binggeli, B.\ 2006, \aj, 132, 497 
\bibitem[Lisker et al.(2006b)]{Lisker2006_2} Lisker, T., et al.\ 2006, \aj, 132, 2432
\bibitem[Marleau et al.(2006)]{Marleau2006} Marleau, F.~R. et al.\ 2006, \aj, 646, 929
\bibitem[Marleau et al.(2009)]{Marleau2009} Marleau, F.~R., Noriega-Crespo, A. \& Misselt, K.\ 2009, AAS, 214, 419.04,
Bulletin of the American Astronomical Society, 41, 688

\bibitem[Mei et al.(2007)]{2007ApJ...655..144M} Mei, S., et al.\ 2007, 
\apj, 655, 144 

\bibitem[Michielsen et al.(2004)]{2004ANS...325..122M} Michielsen, D., de 
Rijcke, S., 
\& Dejonghe, H.\ 2004, Astronomische Nachrichten Supplement, 325, 122 

\bibitem[Pedraz et al.(2002)]{Pedraz2002} Pedraz, S. \ 2002, MNRAS, 332, L59

\bibitem[Pilbratt et al.(2010)]{Herschel} Pilbratt, G., et al.\ 2010, A\&A,
this issue

\bibitem[Poglitsch et al.(2010)]{PACS} Poglitsch, A., et al.\ 2010, A\&A,
  this issue

\bibitem[Roediger 
\& Hensler(2005)]{2005A&A...433..875R} Roediger, E., \& Hensler, G.\ 2005, \aap, 433, 875 

\bibitem[Simien et al.(2002)]{Simien2002} Simien, F. \& Prugniel, P.\ 2002, A\&A, 384, 371

\bibitem[Thomas et 
al.(2006)]{2006A&A...445L..19T} Thomas, D., Brimioulle, F., Bender, R., Hopp, U., Greggio, L., Maraston, C., \& Saglia, R.~P.\ 2006, \aap, 445, L19 

\bibitem[Toloba et al.(2009)]{Toloba2009} Toloba, E., et al.\ 
2009, \apjl, 707, L17 

\bibitem[Valcke et al.(2008)]{Valcke2008} Valcke, S., de Rijcke, 
S., \& Dejonghe, H.\ 2008, \mnras, 389, 1111

\bibitem[van Zee et al.(2004)]{VanZee2004} van Zee, L., Skillman, E.~D., \& Haynes, M.~P.\ 2004, \aj, 128, 121

\bibitem[Vidal \& Baes(2007)]{2007BaltA..16..101V} Vidal, E., \& Baes, M.\ 2007, Baltic Astronomy, 16, 101

\bibitem[Young 
\& Lo(1997)]{1997ApJ...476..127Y} Young, L.~M., \& Lo, K.~Y.\ 1997, \apj, 476, 127 

\end{thebibliography}
\end{document}